# The Ecological System of Innovation: A New Architectural Framework for a Functional Evidence-Based Platform for Science and Innovation Policy

## Robert M. Yawson

Center for Science, Technology, and Public Policy; Hubert H. Humphrey Institute, University of Minnesota, Humphrey Center, 301-19th Ave. South; Minneapolis, MN 55455, USA

E-mail: yawso003@umn.edu

**Abstract:** Models on innovation, for the most part, do not include a comprehensive and end-to-end view. Most innovation policy attention seems to be focused on the capacity to innovate and on input factors such as R&D investment, scientific institutions, human resources and capital. Such inputs frequently serve as proxies for innovativeness and are correlated with intermediate outputs such as patent counts and outcomes such as GDP per capita. While this kind of analysis is generally indicative of innovative behaviour, it is less useful in terms of discriminating causality and what drives successful strategy or public policy interventions. This situation has led to the developing of new frameworks for the innovation system led by National Science and Technology Policy Centres across the globe. These new models of innovation are variously referred to as the National Innovation Ecosystem. There is, however, a fundamental question that needs to be answered: what elements should an innovation policy include, and how should such policies be implemented? This paper attempts to answer this question.

## Summary

It is of no doubt that emerging techniques in the evaluation of R&D and measurement of innovation are finding a broad agreement across nations and regional groupings. However, it is very important for the science policy community across nations to be consistent with the taxonomy used. This paper focuses on the prospective tools, models and data used in this measurement and their importance to the development of a functional evidence-based platform for science and innovation policy. The various models of the National Innovation Ecosystem developed are reviewed and the gaps within these models are identified. The paper then defines a new architectural framework for a National Innovation Policy: The Ecological System of Innovation. In developing this framework the paper uses a hybrid of models, methodologies and concepts including, the Delphi Method, Balanced Scorecard, the Quadruple Helix Theory, and the Analytic Hierarchy Process. The Ecological System of Innovation framework is then used to address the fundamental challenges to the development of a functional evidence-based platform for science and innovation policy.

## Introduction

Researchers have successfully explored the definitions of innovation, innovation theories, the rationale of government interventions on innovation, innovation policy instruments, and the relationships between new technologies, emerging markets, innovative services, and economic growth. However, there are different conceptions regarding what constitutes the core elements of a national innovation system and different scholars draw the borderlines of the system differently.

Innovation is the combination of knowledge that result in new products, processes, input and output markets, or organizations[1] which include not only technical innovations, but also organizational and managerial innovations, new markets, new sources of supply, financial innovations, and new combinations. Innovation is a critical factor in enhancing a nation's competitiveness. National governments have pursued planning in innovation policies to improve their nation's growth.

Industrial innovation includes technical design, manufacturing, management and commercial activities involved in the marketing of a new or improved product or the 1st commercial use of a new or improved process or equipment[2]. Most economic functions in a modern society are best fulfilled by the market mechanism and capitalist firms. However, sometimes there are reasons to complement – or correct – the market and its capitalist actors through public intervention, in such areas as law, education, environment, infrastructure, research, social security, and income distribution. In some of these fields, there is no market mechanism at all, and the functions are fulfilled through the use of other mechanisms, like regulation [3].

Government intervention is therefore needed at certain specific stages of the innovation process. The role of government in innovation is not monolithic. From supporting basic research to building infrastructures and establishing regulations, government policy-makers can define industries and affect fortunes of individual firms. Meanwhile, governments exert a strong influence on the innovation process, via the financing and steering of public organizations that are directly involved in knowledge generation and diffusion like universities and public labs, and through the provision of financial and regulatory incentives. In effect, a national government may play a role in the process of innovation, because of:

1. The public nature of the knowledge that underpins innovation
2. The uncertainty that often plagues the process of innovation
3. The need for certain kinds of complementary assets
4. The nature of certain technologies
5. The type of R&D and funding prioritization
6. Social-wellbeing of citizens and the nation's competitiveness
7. Plain Politics[4]

Edquist and Hommen [5] defined science, technology and innovation (STI) policy in the narrow sense as specific parts of what could be labeled more broadly as 'innovation policy'. Science policy is the most supply-side-oriented of the policies, and least direct. Technology policy is the most difficult to define, because technological research varies significantly within the continuum, from relatively mono-disciplinary scientific research to multi- , inter- , and trans-disciplinary commercial innovation. However, innovation policy that is oriented toward appropriate new product ideas, production processes, and marketing concepts can produce competitive advantages.[6]

The search for appropriate policy tools is not easy. Macro measures are not effective; thus, proposals like a general R&D tax credit are pointless. Policies must be designed to influence particular socioeconomic sectors and activities. The aims and governmental expectations of public policies towards innovation are many and varied, as are the policies themselves and the tools designed to meet policy aims. In this regard, the key policy problem is to address the fundamental challenges to the development of a functional evidence-based platform for science, technology and innovation policy. To address this challenge is to understand the social science of the STI policy. "…the nascent field of the social science of science policy needs to grow up, and quickly, to provide a basis for understanding the enormously complex dynamic of today's global, technology-based society"[7] There is no straight forward answer to the question: what elements should an innovation policy include, and how should such policies be implemented? This paper therefore seeks to answer these questions by defining a new architectural framework for a national innovation policy: the ecological system of innovation (ESI). The ESI will then be used to address the fundamental challenges to the development of a functional evidence-based platform for science and innovation policy. These challenges include[8]:

1. Full systems approach to mapping science, technology and innovation.
2. Portfolio Models of investment in science and technology
3. Behavioral and Dynamic Models of the relationship between scientific discovery and policy decisions
4. Mapping and Cyber Tools linking the evolving taxonomy of science and engineering to policy decision-making
5. Full accounting of intangible assets and international workforce flows, and their contribution to science and technology outcomes.
6. Real-time evaluative and decision-making tools for assessing public sector investments in fundamental S&T on economic growth and social well-being.
7. Measures of spillover effects between scientific discovery and technological innovation, particularly among universities, firms and government labs
8. Evaluative measures of disciplinary cultures on transformative work.
9. Computational models of creativity
10. Evaluative approaches to measuring diversity and its impact on S&T developments.
11. Understanding of human and shifting social dynamics, and the role of S&T in stimulating growth and development.
12. Organizational infrastructure for the collection, collation and interpretation of evidence.

**Background**

Models on innovation, for the most part, do not include a comprehensive and end-to-end view. Most innovation policy attention seems to be focused on the capacity to innovate and on input factors such as R&D investment, scientific institutions, human resources and capital. Such inputs frequently serve as proxies for innovativeness and are correlated with intermediate outputs such as patent counts and outcomes such as GDP per capita. While this kind of analysis is generally indicative of innovative behavior, it is less useful in terms of discriminating causality and what drives successful strategy or public policy interventions. This situation has led to the developing of new frameworks for the innovation system led by National Science and Technology Policy Centers across the globe. These new models of innovation are variously referred to as the National Innovation Ecosystem.

Earlier models created for the measurement of innovation are disparate, mostly linear in nature and neglecting the interrelationships between the various indicators and metrics. Statistics on science are often framed within an input–output framework: inputs are invested into research activities that produce outputs. This framework is a pure accounting framework based on the anticipated economic benefits of science[9]. Most existing methodologies for measuring innovation are motivated by research and practice in domains of accounting, economics, human resource accounting, intellectual property, and, real options, among others. Prior reviews of such models have focused at the firm level analysis with an accounting, economic, or strategic lens. [10] Most of these models have not been directly applied for assessment of national innovation systems.

Many empirical research studies and institutional policy frameworks do however relate to the key elements of these models in their conceptualization. However, the main gap identified in the literature with these new national innovation policy portfolios (popularly referred to as National Innovation Ecosystem) is the linear nature of presentation just as the previous models of innovation.

An ecosystem is not linear, it is a web of interrelationships, different systems, niches and pathways coming together to sustain life. An innovation ecosystem framework should not be linear but rather a web, an interlocking systems and pathways helping to sustain and formulate a functional evidence-based policy making. An ecological system of innovation can be constructed at a number of levels of abstraction and detail—from an individual technology project, to the enterprise, to the industry sector, to the national, regional and even global level.

The new architectural framework for a national innovation policy proposed in this paper extends the traditional linear chain model to the innovation process and enlarges it to incorporate all aspects of society, including academia, government, industry and the public, thus creating a comprehensive National Ecological System of Innovation. Despite a national outlook, the framework retains focus on the organizational level and metrics and tools for measurement at that level.

In order to understand the innovation process it is necessary to focus upon interaction and relationships. Organizations, knowledge institutions and people do seldom innovate alone and innovation emanates from cumulative processes of interactive learning and searching. This implies that the system needs to be characterized simultaneously with reference to its elements and to the relationships between those elements. The relationships may be seen as carriers of knowledge and the interaction as processes where new knowledge is produced and diffused.[11] The key issue facing many organizations relates to how they can foster effective innovation using organizational supporting mechanisms. Failure to innovate is likely to result in reduced competitiveness. There is the need for theoretical integration to link organizational context with innovation and to consider strategic orientation as an important action parameter and the general framework for decisions about innovation and change.[12] It is the contention of this author that to address the roles that organizations and organizational change play in knowledge creation and in the national systems of innovation, it must be understood that:

1. Organizations play the most important role in the innovation system.
2. Organizations innovate in an interaction with other organizations and with knowledge infrastructure.
3. Organizations' mode of innovation and learning reflect national innovation systems
4. Organizations belonging to different sectors contribute differently to innovation processes.

In this author's opinion there are three main domains that are foundational to addressing the roles that organizations and organizational change play in scientific discovery processes and innovation:
1. organizational learning,
2. knowledge creation and transfer or conversion
3. Strategic change implementation, change management, specifically, process innovation.

Organizational level learning refers to a collective, group or whole organization, "thinking" and behaving differently as a result of a change process. The knowledge creating and transfer process refers to those activities, both human and technology based, that change cognition. The change management is an organizational initiative designed to achieve significant improvements in performance by changing relationships between people, technology, organizational structure, and information, and which typically begins with a strategic change to which the top management team is committed.

Without a basic understanding of the combination of organizational and inter-organizational learning it is impossible to establish the link from innovation to economic growth. To put it briefly the focus should be much more on people and competence and upon how the relationships and interactions between people promote learning. This is especially important in the current era of the globalizing learning economy where the key to success for individuals, firms, regions and national systems is rapid learning. [13] To understand how learning takes place within organizations as well as in the interaction between organizations is a key to understand how a system of innovation works.

**Ecological System of Innovation**

The new architectural framework for a national innovation policy was initiated by reviewing the different models and innovation portfolios developed or being developed by different countries and organizations. This was made relatively easy by reviewing two conference proceedings:

1. KISTEP-WREN Workshop/International Symposium on National Models for Public R&D Evaluation: In Search of Best Practices and Collaborative Opportunities held in May 30-31, 2005 and was Organized and Hosted by Korea Institute of S&T Evaluation and Planning (KISTEP); Co-organized by Washington Research Evaluation Network (WREN), and the EU Commission. The conference brought together national leaders in R&D from the USA, EU, Japan, Korea, India, China and other countries[14]

2. The International Symposium on Innovation Policy and Evaluation was held in Japan from November 19 – 20, 2007 in conjunction with the G8 Working Group on Research Assessment in Tokyo.[15] In addition to the delegates of the G8 countries numerous eminent researchers in the fields of R&I policy were invited from abroad. Key figures from such organizations as PRIME and WREN, along with additional members representing Korea, China, and Japan participated. In this symposium "innovation" was defined as "the introduction of something new" to cover both socio-economic and *scientific* innovation. The aim of the symposium was to review the global frontiers of innovation policy and evaluation in practice, while at the same time laying a solid foundation for a new global network of R&I policy studies and evaluation.

These two conferences/workshops were selected because they brought together more or less "mouthpieces" of governments: The highest level of national representatives. They provided credible and authentic sources of what the respective countries are doing in the area innovation policy portfolios.

From the synthesis of the reports and presentations, it is of no doubt that emerging models, datasets, metrics and methodologies used in the evaluation of R&D and measurement of innovation are finding a broad agreement across nations and regional groupings. However, it is very important for the science policy community across nations and regions to be consistent with the taxonomy used. The main gap identified in analyzing the various models of the National Innovation Ecosystem is the linear nature of presentation just as the previous models of innovation system.

The paper therefore defines a new architectural framework for a National Innovation Policy: The Ecological System of Innovation. In developing this framework the paper uses a hybrid of models, methodologies and concepts including, the Delphi Method, the DEMATEL Method, the Balanced Scorecard (BSC), Systems Dynamics Modeling, the Quadruple Helix Theory (QHT), and the Analytic Hierarchy Process (AHP). The architectural framework (fig. 1) consists of four main phases: (1) defining or establishing the national innovation requirements using the Delphi Method; (2)Gap Identification and solution space, using DEMATEL and all the prospective tools and models including Scientometrics, Complex Adaptive Systems, etc;

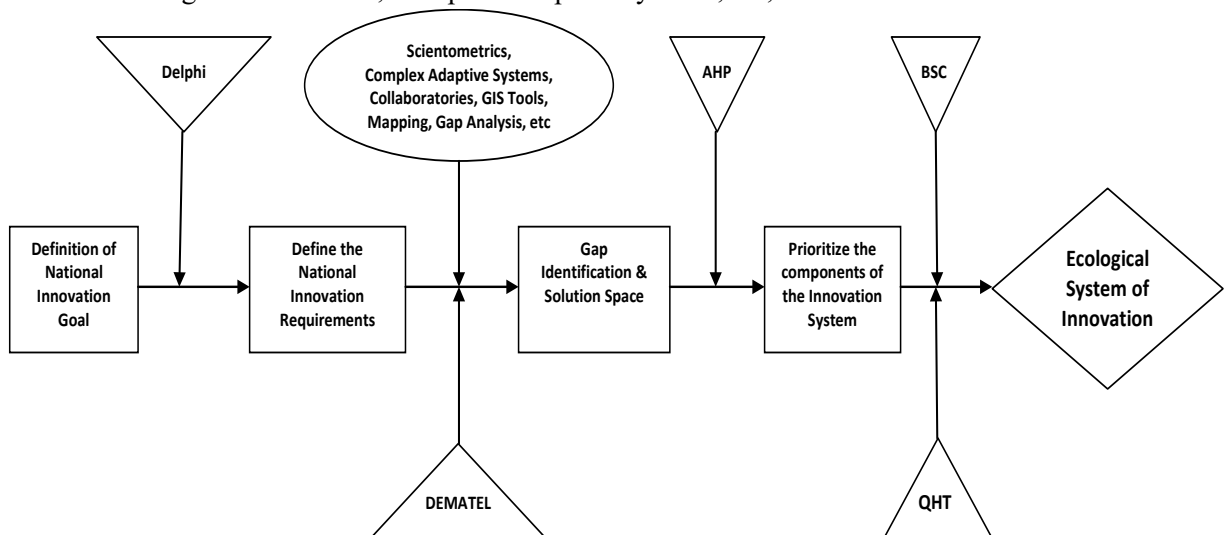

*Fig. 1: The Architectural Framework for defining an Ecological System of Innovation. ©R. M. Yawson 2008*

(3) Correlating the national innovation requirements and policy tools and, thus ranking the priorities of the innovation policy tools with AHP; and finally (4) Constructing the Ecological System of Innovation using QHT and the BSC and Systems Dynamics Modeling.

### Delphi Method

The Delphi method originated in a series of studies conducted by the RAND Corporation in the 1950s.[16] The objective was to develop a technique to obtain the most reliable consensus from a group of experts. While researchers have developed variations of the method since its introduction, including the very popular Expert Elicitation Methods, Linstone and Turoff[17] captured its common characteristics in the following description: Delphi may be characterized as a method for structuring a group of communication process; so the process is effective in allowing a group of individuals, as a whole to deal with a complex problem. To accomplish this 'structured communication', certain aspects should be provided: some feedback of individual contributions of information and knowledge; some assessment of the group judgment or viewpoint; some opportunity for individuals to revise their views; and some degree of anonymity for individual responses. The Delphi technique enables a large group of experts to be surveyed cheaply using self-administered questionnaire. The Delphi method proceeds in a series of communication rounds, as follows:

<u>Round 1</u>: Either the relevant individuals are invited to provide opinions on a specific matter, based on their knowledge and experience, or the team undertaking the Delphi expresses opinions on a specific matter and selects suitable experts to participate in subsequent questionnaire rounds; these opinions are grouped together under a limited number of headings, and statements are drafted for circulation to all participants through a questionnaire.

<u>Round 2:</u> Participants rank their agreement with each statement in the questionnaire; the rankings then are summarized and included in a repeat version of the questionnaire.

<u>Round 3:</u> Participants re-rank their agreement with each statement in the questionnaire, and have the opportunity to change their score, in view of the group's response; the re-rankings are summarized and assessed for the degree of consensus: if an acceptable degree of consensus is obtained, the process may cease, with the final results then fed back to the participants; if not, this third round is repeated. Therefore in developing a National Innovation Policy, the first step after goal definition (which is more of a political decision), is to define the national innovation requirements and the Delphi Method is one of the most appropriate methods to be used. Participants should include experts from academia, government, industry and the public.

### *Prospective Tools and Models*

There are several analytical tools being developed for the collection, collation and interpretation of STI metrics and indicators for innovation. After the national innovation requirements are defined, there is the need to undertake gap Identification and then find a solution space. A highly functioning innovation system is expected to have the following attributes and benchmarks: Competition for Resources (Money, Ideas, People, Facilities); An open market place for ideas (Patents, Papers, Copyrights, IP); Resources sufficient for system growth (People, Money, Land, Energy); and Checks & Balances (Transparency, Multiple Funding Sources, External Review).[18] Gap analysis naturally flows from benchmarking and other assessments; and in STI policy the assessments include STI metrics, scientometrics and all the tools used in evaluating R&D and measuring innovation. Once the national innovation requirements have been defined it is possible to compare the results from the Delphi Method with the level of performance of existing innovation indicators. This comparison becomes the gap analysis. Gap analysis is a formal study of what is being done currently and what needs to be done in the future. It provides a foundation for measuring tangible and intangible indicators required for the ecological system of innovation.

There are several other tools that can be used at this stage. Some of these tools are frontier and still undergoing refinements in their use and applicability. These tools are needed for the collection of STI metrics and creation of the relevant Science and Engineering Indicators (SEI). SEI are first and foremost a volume of record comprising the major high-quality quantitative data on the national and international science and engineering enterprise.[19] SEI should employ a variety of presentational styles—tables, figures, narrative text, bulleted text, Web-based links, highlights, introductions, conclusions, reference lists—to make the data accessible to readers with different information needs and different information-processing preferences. [20] SEI should include increased participation in the sciences, increased numbers of people with advanced qualifications, enhanced contribution by research to economic and social development, transformational change in the quality and quantity of research, increased output of economically relevant knowledge, increased trans-national research activity, and greater coherence and exploitation of synergies nationally and internationally. Quantitative material on the state-of-the art in science and technology has seen a strong increase since the second half of the last century. Major federal agencies of the United States, UNESCO, OECD, and the European Union Commission are main examples of organizations that collect systematically data on the development of science, technology and innovation.

New ideas, on the one hand, and new R&D professionals, on the other hand, are the two major outputs of academic science. The latter are relatively easy to measure, whereas the output of new ideas is more difficult to grasp in an objective, and preferably quantitative, way. Tools like scientometrics can be used in this regard. Scientometrics is the science of measuring and analyzing science. In practice, scientometrics is often done using bibliometrics i.e. measurement of (scientific) publications. [21]

Collaboratories are another way data and Innovation indicators can be obtained nationally and internationally. The term collaboratory was coined from two words collaborate and laboratory.[22] The term was defined by William Wulf in 1989[23], as a "center without walls, in which the nation's researchers can perform their research without regard to physical location, interacting with colleagues, accessing instrumentation, sharing data and computational resources, and accessing information in digital libraries". The definition was later refined[24] to "a system which combines the interests of the scientific community at large with those of the computer science and engineering community to create integrated, tool-oriented computing and communication systems to support scientific collaboration". There are several other prospective tools and models that can be used for the stage two of the architectural framework, but discussing them is beyond the scope of this paper.

*DEMATEL Method*
Decision Making Trial and Evaluation Laboratory (DEMATEL) may also be introduced at the stage 2. The DEMATEL method was developed by Battelle Geneva Institute:[25] (1) to analyze complex 'world problems' dealing mainly with interactive man-model techniques; and (2) to evaluate qualitative and factor linked aspects of societal problems. The applicability of the method is widespread, ranging from industrial planning and decision-making to urban planning and design, regional environmental assessment, analysis of world problems and so forth. The DEMATEL method is based on graph theory, enabling the planning and solving problems visually, so that multiple criteria may be divided into cause and effect group, in order to better understand causal relationships. This methodology was introduced at the stage 2 of the framework not only to help identify gaps and create solution space, but also to be a tool to evaluate the usefulness of all the methodologies, models and tools to be used at that stage.

*Analytic Hierarchy Process (AHP)*
Use of relative measures in decision-making has been demonstrated to be more valuable than decisions taken based on absolute measures, especially when considering tangibles and intangibles together.[26] Therefore, it is advisable to use relative scoring methods when attempting to assess the relative importance (or relative dominance or relative preference) of factors and actors that contribute to taking a decision. The AHP developed by Saaty[27] for complex decision problems is a technique used for data analysis to develop a prioritization of relative importance of model factors. It is a decision-making framework that uses a hierarchical structure to describe a management problem. AHP paired comparisons rank all items at each level with respect to their relative importance with each other, and then convert level-specific local priorities into broader level decision priorities.[28] An AHP hierarchy is a structured means of describing the problem at hand. It consists of an overall *goal*, a group of options or *alternatives* for reaching the goal, and a group of factors or *criteria* that relate the alternatives to the goal. In most cases the criteria are further broken down into sub-criteria, sub-sub-criteria, and so on, in as many levels as the problem requires. This methodology is therefore ideal for the stage three of the framework: Correlating the national innovation requirements and policy tools and, thus ranking the priorities of the innovation policy tools is a very critical stage in the development of the Ecological System of Innovation.

### Quadruple Helix Theory (QHT)

In recent years, social scientists and others have shown an increased level of interest in analyzing the enterprise of science. Since the 1950s, developments in science have stimulated major innovations in nuclear technology, information and communication technology, biotechnology and nanotechnology. In addition to stimulating innovation and convergence, these technologies have generated also a pressing social need to understand the risks and benefits that flow from such innovations, and have brought into question the enterprise of science and the adequacy of regulations. The triple helix of state, university and industry is missing an essential fourth helix, the public. Advances in biotechnology and nanotechnology are jeopardized by the virtual absence of this helix. The proposed framework of National Ecological System of Innovation incorporates this fourth helix. Disciplinarity is no longer the dominant system for creating and organizing knowledge. Knowledge creation is now trans-disciplinary, more reflexive, non-linear, complex and hybridized. Furthermore, inclusion of the fourth helix becomes critical since scientific knowledge is increasingly evaluated by its social robustness and inclusivity. Public interest is important in this regard. The fourth helix highlights new discoveries and innovations that improve social welfare e.g. eco-innovation. It helps to create linkages between science, scientists and education strategies. The QHT is made part of the final stage of the framework because of the need to include the public in the innovation system.

### The Balance Scorecard (BSC)

The final stage in the architectural framework for the National Ecological System of Innovation is the use of the BSC methodology. The BSC was first developed by Kaplan and Norton in 1992.[29] Leading organizations agree on the need for a structured methodology for using performance measurement information to help set agreed-upon performance goals, allocate and prioritize resources, confirm or change current policy or program directions to meet those goals, and report on the success in meeting those goals. BSC methodology is used for deploying strategic direction, communicating expectations, and measuring progress towards agreed-to objectives. The balanced scorecard is a conceptual framework for translating an organization's strategic objectives into a set of performance indicators distributed among four perspectives: Financial, Customer, Internal Business Processes, and Learning and Growth. Some indicators are maintained to measure an organization's progress toward achieving its vision; other indicators are maintained to measure the long term drivers of success. Through the balanced scorecard, an organization monitors both its current performance (finance, customer satisfaction, and business process results) and its efforts to improve processes, motivate and educate employees, and enhance information systems—its ability to learn and improve. The BSC complements information provided by other tools with its process-based focus on how specific actions relate to organizational performance outcomes.

Given that this model is particularly conducive for relating the strategic vision to core competencies and related success factors for organizational success, it provides one possible basis for developing an action blueprint for the National Ecological System of Innovation. Using the BSC to develop the National Ecological System of Innovation is novel. The first step in the architectural framework is the definition of the national goal; which is to "develop a functional Evidence-Based Platform for Science and Innovation Policy". The BSC translates an organization's vision into a set of performance objectives distributed among four perspectives. In this framework the four perspectives are those adapted from the QHT: academic perspectives, government perspectives, industry perspectives and public perspectives. Each objective within a perspective should be supported by at least one measure that will indicate an organization's performance against that objective.

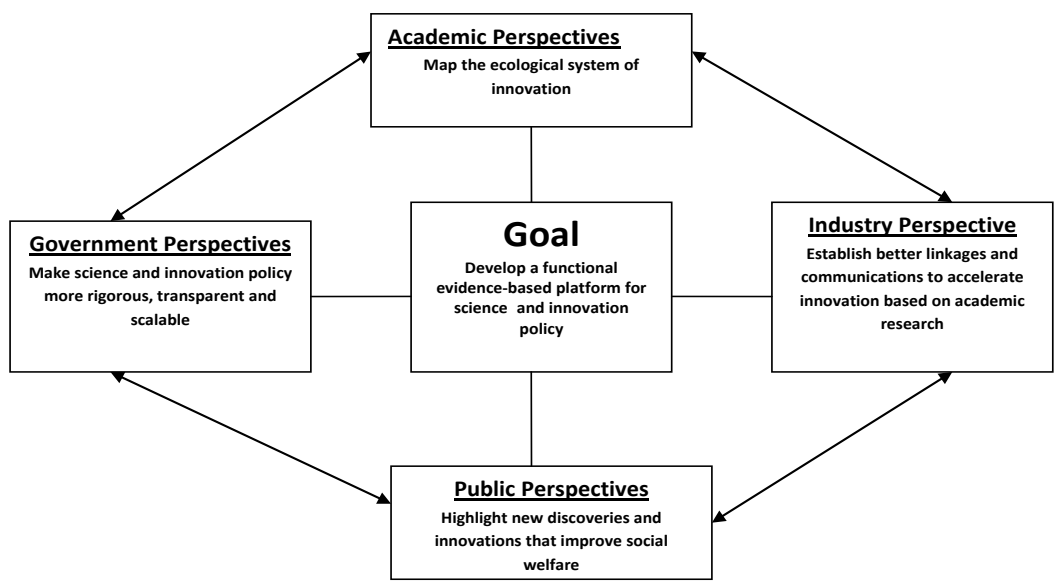

**Fig 2: Science and Innovation Policy Scorecard**

One of the most important features of this methodology is that it is dynamic system, i.e., it allows for enhancements over time, in light of changing circumstances. While some measurement concepts may seem timeless, the ever-changing character of STI policy dictates that maintenance of a current measurement model is a priority. The model may need to be updated periodically to reflect changes in the National Ecological System of innovation. Also, there may be a need to discard measures that have not proved useful, or to modify existing core measures to enhance their utility. However, any system revisions will be made on a selective basis to ensure that the BSC permits national organizations to gauge performance progress against a consistent baseline, and to ascertain and analyze meaningful trends. The Balanced Scorecard (BSC) complements information provided by other tools with its process-based focus on how specific actions relate to organizational performance outcomes.

The strength of the Ecological System of Innovation Model lies not only in the consideration of the four perspectives, but also the way in which these perspectives interrelate, and contribute to the national goal. The mapping of objectives – looking at cause and effect relationships – visualises how the objectives are linked. Mapping has two purposes at this point:

> ➢ Firstly, as a tool to help strategize and prioritise areas for development.

> Secondly, once the ecological system of innovation has been established, mapping will potentially help identify blockages, enabling corrective action to be taken.

Key Performance Indicators (KPIs) have a crucial role to play as measures of the success of each objective, and as indicators of the likelihood of the linked objective being met. Having established and tested the key linkages between objectives, it may be necessary to review the KPIs, to see whether or not that effectively fulfils this function. If not, they may need to be adjusted or added to, or it may be considered appropriate to develop some extra KPIs to look at the interface between one or more objectives.

**Table 1 showing the academic perspectives of the Scorecard**

| *Goal:* | Map the Ecological System of Innovation |
|---|---|
| *Objectives* | *Key Performance Indicators and tools* |
| Develop a new taxonomy of science, technology and innovation | Bibliometric, scientometric and visualization tools to mine patent, publication and CV data to establish linkages between known disciplines and to establish births and deaths of fields |
| Establish the theory and practice on organizational structure that facilitates discovery and innovation | Creativity: Computational models of the creativity process. Organizations: Business-2-Business, industry-university partnerships; collaboratories/virtual organizations; interdisciplinary, multidisciplinary, transdisciplinary teams |
| Gather data and create metrics of the ESI | Workforce data: S&E, postdocs, graduate students, undergraduates; foreign born and domestic. Expenditure data: R&D-federal, state, firms, academic. Output data: patents, peer-reviewed publications, intangible assets. Innovation metrics |
| Develop new methodologies to reduce uncertainty in prospective policies. | Develop new measures for assessing basic research and evaluating novel ideas/proposals |
| Develop a theoretical and analytical model/frame work for linking input metrics to scientific achievement or societal benefit | |

**Table 2 showing the government perspectives of the Scorecard**

| Government Perspective |
|---|
| *Goal*: Make Science and innovation policy more rigorous, transparent and scalable |
| *Objectives* |
| Develop econometric models that utilize network analysis and a systems approach to stimulate policy scenarios, creating a portfolio of scientific investments |
| Create a roadmap for funding agencies to use as they implement new datasets, metrics, tools, and models in their decision-making processes |
| Include risk-analysis & probability tools to determine the capacity of alternative science policies for achieving desirable outcomes |
| Develop/Use benchmarking tools that present realistic scenarios for ascertaining comparative and competitive advantages in the international arena |
| Adopt, modify (if necessary) and implement best practices from industry relating to funding/investment decision |
| Assess spillover effects of funding decisions in basic sciences |
| Develop a cost benefit analysis procedure that enables policy makers at various levels of decision-making to prioritize investments in science, technology and innovation |

**Table 3 showing the Industry perspectives of the Scorecard**

| Industry perspective |
|---|
| *Goal*: Establish better linkages and  communications to accelerate innovations based on academic research |
| *Objectives:* |
| Bridge "valley of death" |
| Create links between public and industry innovation to study the ultimate effect. Gathering matrix of public reaction and adoption of innovative products should be useful in designing pathways for future innovation in product/processes. |
| Utilize tools for better assessment of uncertain technologies and making investment decisions in R&D. |

**Table 4 showing the Industry perspectives of the Scorecard**

| Public perspective |
|---|
| *Goal*: Highlight new discoveries and innovations that improve social welfare e.g. eco-innovation |
| *Objectives*: |
| Create a sectoral report on research output as they relate to or address social issues. |
| Create linkages between science, scientist and education strategies |

**System Dynamics Modeling of the BSC for the ESI**

The reason for introducing SDM into the BSC discussion is mainly to be able to address any time lag issues related with dynamic environments. In addition, it demonstrates the benefit of using SDM for a concept like BSC, and to shed some light on the formulation of the timing aspects pertaining to the cause-and-effect relations between BSC means and measures. System dynamics is a perspective and a set of conceptual tools that enables the understanding of the structure and dynamics of complex systems. It is also a rigorous modeling method that enables the building of formal computer simulations of complex systems and uses them to design more effective policies and organizations. The model may be used as the first step in quantifying the cause-and-effect relationships of an integrated ESI model. Using the system dynamics model provides added insight in the BSC.

**Conclusion**

The architectural framework developed answers our initial question on what elements should an innovation policy include, and how should such policies be implemented? The various stages in the framework describe what elements the innovation policy should include. The development of the Ecological System of Innovation answers the how should the policies be implemented.

The ESI is more than a collection of measurement indicators as all the measures are linked through a chain of cause-and-effect that culminates into strategic success. The cause-and-effect hypothesis is fundamental to understanding the metrics that the balanced scorecard prescribes and how they relate to strategic success. Therefore, the policy analysts need to continuously assess if the chosen policies are correctly implemented (as determined by specific indicators) and then ensure that the assumptions made about cause and effect relationships are evident in practice. If the specific value-added performance outcomes are not achieved, the causal links need to be reassessed to ensure that the constructs as well as their relationships are valid. This is the main reason for the use of systems dynamics modeling in this new architectural framework.

The process of developing the ESI starts with the national innovation goal that is interpreted through the four perspectives: Academic, Government, Industry, and Public. The goal is translated into competencies relevant to each of the four perspectives along with an assessment of critical success factors and specific indicators that represent the inputs, processes, outputs, and outcomes for each of the four perspectives. ESI recognizes that innovation by creative citizens, presence of a learning and knowledge sharing culture, and formal and informal learning opportunities underpin the success of the innovation goal and strategy. Learning and growth are fostered through knowledge management activities and initiatives such as strategic recruiting, hiring, training, team development, document management, collaborative communication systems, knowledge and skills audits of employees, knowledge base developments, and fostering of communities of interest within organizations.

Review of research and policy literatures on innovation indicates growing interest in knowledge economies and knowledge societies that can promote holistic social, cultural, economic development and well being of citizens. This study reviewed the concepts of National Innovation Ecosystem; compared and contrasted most popular measurement and management models and methodologies; critiqued the current models and indicators in use; and proposed an actionable blue-print – an architectural framework – for developing public sector capacities in innovation policy-making.

The following discussion summarizes the framework developed and provides directions for future research and development for further improvements in the measurement models. The key observations and recommendations based upon the review, analysis, and development of measurement methodology and frameworks for national innovation system are listed below. These observations represent critical 'thinking points' that can help define the contours and trajectory of the emerging innovation systems. These observations define key issues that need to be further developed in terms of specific research agendas and policy applications.

1. Significant progress has been achieved in terms of development of measurement models of innovation for analysis at the firm level. There has been some progress in developing similar models for assessment of national innovation systems and for enabling innovation policy-making. Drawing upon a comprehensive review of the research and practice literatures as well as national policy documents, this study has attempted to fill this void.
2. Most existing models for measuring innovation and innovation policy suffer from a critical 'disconnect.' Their reliance upon the inputs as valid proxies of performance outcomes raises a very critical issue: if they indeed measure what they attempt to measure. Investments in public sector projects cannot be considered proxies for performance outcomes. Similarly, investments in developing structural artifacts and processes for 'getting things done' are not valid proxies for 'things that need to be done.' Also, what is actually done or delivered has to justify as value creation based upon prior expectations to conclude that performance outcomes have indeed been achieved. This paper has proposed a model for linking inputs-processes-outputs-outcomes for measuring national innovation and enabling public sector competencies for such measurement. Future research and development is needed for further improving the predictive validity of the measurement models and related indices and indicators.

3.  While large number of empirical studies have been conducted around the world on innovation most of these studies have followed the accounting and economic perspectives. These disciplines share common criteria about evaluation of 'assets' and 'capital' even though conflicting national and regional accounting standards make cross-national comparison a challenge. There has been growing recognition about developing complementary perspectives from disciplines such as sociology and psychology that can provide richer assessment of social and behavioral issues. An encouraging development is the recent developmental studies by organizations such as the OECD that share the concern about better indices and indicators related to innovation. Sociological and behavioral issues such as social influence, persuasion, self-determination, commitment, and, intrinsic motivation are directly relevant to the content and quality of performance outcomes wherever human agents are involved. [**30**] Better understanding of such intangibles' is needed to enrich and refine the constructs of the public perspective of the ESI model.

4.  While significant development of existing measurement frameworks and methodologies has occurred in the past years, fundamental theoretical concerns loom large. For instance, in absence of a generally accepted theory of innovation policy, how much confidence can we place in measurement models of innovation. Growth of multi-disciplinary, trans-disciplinary and inter-disciplinary theoretical foundations that can integrate the concerns raised in this study is recommended for analysis of complex constructs that defy the bounded logic of specific disciplines. The current stream of research that has attempted to develop the sociological and behavioral understanding of science policy seems relevant in this regard.